\documentclass[12pt,preprint]{aastex}

\shorttitle{GRB 050502B X-ray Flare}
\shortauthors{Falcone et al.}

\begin{document}


\title{The Giant X-ray Flare of GRB 050502B: Evidence for Late-Time Internal Engine Activity}


\author{
A.~D.~Falcone\altaffilmark{1,2},
D.~N.~Burrows\altaffilmark{1}, 
D.~Lazzati\altaffilmark{3},
S.~Campana\altaffilmark{4},
S.~Kobayashi\altaffilmark{1,5},
B.~Zhang\altaffilmark{6},
P.~M\'{e}sz\'{a}ros\altaffilmark{1,7},
K.~L.~Page\altaffilmark{8},
J.~A.~Kennea\altaffilmark{1},
P.~Romano\altaffilmark{4},
C.~Pagani\altaffilmark{1,4},
L.~Angelini\altaffilmark{9,10},
A.~P.~Beardmore\altaffilmark{8},
M.~Capalbi\altaffilmark{11},
G.~Chincarini\altaffilmark{4,12},
G.~Cusumano\altaffilmark{13},
P.~Giommi\altaffilmark{11}, 
M.~R.~Goad\altaffilmark{8},
O.~Godet\altaffilmark{8},
D.~Grupe\altaffilmark{1},
J.~E.~Hill\altaffilmark{1,9},
V.~La~Parola\altaffilmark{13},
V.~Mangano\altaffilmark{13},
A.~Moretti\altaffilmark{4},
J.~A.~Nousek\altaffilmark{1}, 
P.~T.~O'Brien\altaffilmark{8},
J.~P.~Osborne\altaffilmark{8},
M.~Perri\altaffilmark{11},
G.~Tagliaferri\altaffilmark{4}, 
A.~A.~Wells\altaffilmark{14},
N.~Gehrels\altaffilmark{9}
}

\altaffiltext{1}{Department of Astronomy \& Astrophysics, Pennsylvania State University, University Park, PA 16802, USA}
\altaffiltext{2}{corresponding author email afalcone@astro.psu.edu}
\altaffiltext{3}{JILA, University of Colorado, Boulder, CO 80309, USA}
\altaffiltext{4}{INAF -- Osservatorio Astronomico di Brera, Merate, Italy}
\altaffiltext{5}{Center for Gravitational Wave Physics, Pennsylvania State University}
\altaffiltext{6}{Department of Physics, University of Nevada, Las Vegas, NV}
\altaffiltext{7}{Department of Physics, Pennsylvania State University, University Park, PA 16802, USA}
\altaffiltext{8}{Department of Physics and Astronomy, University of Leicester, Leicester, UK}
\altaffiltext{9}{NASA/Goddard Space Flight Center, Greenbelt, MD}
\altaffiltext{10}{Universities Space Research Association, Columbia, MD}
\altaffiltext{11}{ASI Science Data Center, Frascati, Italy}
\altaffiltext{12}{Universit\`a degli studi di Milano-Bicocca, Dipartimento di Fisica, Milano, Italy} 
\altaffiltext{13}{INAF- Istituto di Fisica Spazialee Fisica Cosmica sezione di Palermo, Palermo, Italy}
\altaffiltext{14}{Space Research Center, University of Leicester, Leicester, UK}


\begin{abstract}

Until recently, X-ray flares during the afterglow of gamma ray bursts (GRBs) were a rarely detected phenomenon, thus their nature is unclear.  During the afterglow of GRB 050502B, the largest X-ray flare ever recorded rose rapidly above the afterglow lightcurve detected by the {\it{Swift}} X-ray Telescope.  The peak flux of the flare was $>500$ times that of the underlying afterglow, and it occurred at $>12$ minutes after the nominal prompt burst emission.  The fluence of this X-ray flare, (1.0 $\pm$ 0.05) $\times 10^{-6}$ erg cm$^{-2}$ in the 0.2--10.0 keV energy band, exceeded the fluence of the nominal prompt burst.  The spectra during the flare were significantly harder than those measured before and after the flare.  Later in time, there were additional flux increases detected above the underlying afterglow, as well as a break in the afterglow lightcurve.  All evidence presented below, including spectral and particularly timing information during and around the giant flare, suggests that this giant flare was the result of internal dissipation of energy due to late central engine activity, rather than an afterglow-related effect.  We also find that the data are consistent with a second central engine activity episode, in which the ejecta is moving slower than that of the initial episode, causing the giant flare and then proceeding to overtake and refresh the afterglow shock, thus causing additional activity at even later times in the lightcurve.

\end{abstract}





\section{Introduction}

Since its launch on 2004 November 20, {\it{Swift}} \citep{geh04} has provided detailed measurements of numerous GRBs and their afterglows with unprecedented reaction times.  Of the 57 bursts detected by the Burst Alert Telescope (BAT; Barthelmy et al. 2004) as of 2005 August 3, 43 were observed by the narrow field instruments in less than 200 ks (typical reaction time was much less, but occasionally BAT detected a burst that was observationally constrained).  The narrow field instruments include the X-ray telescope (XRT; Burrows et al. 2005) and the Ultraviolet-Optical Telescope (UVOT; Roming et al. 2005).  Of these 43 observations, 42 afterglows were detected by the XRT, and 30 of them received prompt ($< 300$ s) observations with the pointed instruments.  By detecting burst afterglows promptly, and with high sensitivity, the properties of the early afterglow and extended prompt emission can be studied in detail for the first time.  This also facilitates studies of the transition between the prompt emission and the afterglow.

While there are still many unknown factors related to the mechanisms that produce GRB emission, the most commonly accepted model is that of a relativistically expanding fireball with associated internal and external shocks \citep{mes97}.  In this model, internal shocks produce the prompt GRB emission. Observationally, this emission typically has a timescale of $\sim30$ s for long bursts and $\sim$0.3 s for short bursts \citep{mee96}.  The expanding fireball then shocks the ambient material to produce a broadband afterglow that decays quickly (typically as ${\sim}t^{-\alpha}$).  When the Doppler boosting angle of this decelerating fireball exceeds the opening angle of the jet into which it is expanding, then a steepening of the lightcurve (jet break) is also predicted \citep{rho99}.  For a description of the theoretical models of GRB emission and associated observational properties, see \citet{mes02}, \citet{zha04}, \citet{piran05}, and \citet{van00}.

Several authors have suggested reasons to expect continued activity from the internal engine of the GRB after the classical "prompt" emission time frame.  \citet{kat97} considered a model in which a magnetized disk around a central black hole could lead to continued energy release in the form of internal shocks.  The parameters of this energy release would depend on the complex configuration of the magnetic field and the magnetic reconnection dynamics, but time periods as long as days for the delayed emission were predicted.  \citet{kin05} have speculated that episodic accretion processes could explain continued internal engine activity.  These authors expect that fragmentation and subsequent accretion during the collapse of a rapidly rotating stellar core could explain observations of extended prompt emission.  In general, the dominant model of an expanding fireball with internal/external shocks \citep{mes97} allows for continued prompt emission, provided that the internal engine is capable of continuing the energy injection.

A few previous observations have included indications of flaring from GRBs after the nominal prompt emission phase.  \citet{wat03} used XMM-Newton to detect line emission from GRB 030227 nearly 20 hours after the prompt burst.  They inferred continued energy injection at this late time, and concluded that a nearly simultaneous supernova and GRB event would require sporadic power output with a luminosity in excess of $\sim5\times10^{46}$ erg s$^{-1}$.  \citet{pir05} used Beppo-SAX to observe two GRBs with relatively small X-ray flares.  The X-ray flare times for GRB 011121 and GRB 011211 were reported as t=240 s and t=600 s, respectively.  The spectral parameters of these two X-ray flares were consistent with afterglow parameters, and these flares were interpreted as the onset of the afterglow \citep{pir05}.  Two other examples of flaring and/or late timescale emission can be found in \citet{int03} and \citet{gal05}.  Although not a detection of late flares from a particular GRB, the work of \citet{con02}, in which an ensemble of GRBs was analyzed, should also be mentioned.  In this study, 400 long GRBs detected by the Burst and Transient Source Experiment (BATSE) were analyzed together in the form of a summed lightcurve above 20 keV.  Significant emission was found at late times (at least to 1000 s).  There are several possible explanations for this emission that do not require flares, but flares at various times are certainly one possible explanation.

More recently, \citet{bur05b} provided the initial report that two bursts detected by {\it{Swift}} showed strong X-ray flares.  The first of these, XRF 050406, was an X-ray flash with a short, and relatively weak, X-ray flare that peaked 213 s after the nominal prompt emission.  Due to the fast rise/decay, the most natural explanation for this flare is continued internal engine activity at late times (i.e. delayed prompt emission).  A detailed analysis of XRF 050406 is currently in preparation \citep{romano05}.  GRB 050502B, the subject of this paper, was also reported on by \citet{bur05b} since it had a dramatic X-ray flare that peaked 740 s after the nominal prompt emission.  This paper will now explore this event in more detail.

GRB 050502B was detected by the {\it{Swift}}-BAT at 09:25:40 UT on 2005 May 02 \citep{fal05}.  According to \citet{cum05}, the T90 duration for the prompt emission detected by BAT was (17.5 $\pm$ 0.2) s, and the burst had three individual peaks.  The main hard peak had a 6 s duration, was well fit by a power law with photon index 1.6 $\pm$ 0.1, and had a 15--350 keV fluence of (8.0 $\pm$ 1.0) $\times10^{-7}$ erg cm$^{-2}$ \citep{cum05}.  The spacecraft slewed promptly and observations with {\it{Swift}}-XRT and {\it{Swift}}-UVOT began 63 s after the BAT trigger time.  Since the flux was initially low, the XRT Image Mode data did not produce an initial onboard centroid position; however the first pass of data was analyzed on the ground, leading to an XRT position of RA $09^{h}30^{m}10.1^{s}$, Dec $+16^{\circ}59^{m}44.3^{s}$ (J2000), with a 90\% containment uncertainty of 5$\arcsec$ \citep{pag05}.  There was no counterpart found by UVOT, but ground-based optical observations reported by \citet{cen05} revealed a fading afterglow at RA 09:30:10.02, Dec +16:59:48.07 (J2000), which is within 4$\arcsec$ of the reported XRT position.  Following the initial low-flux detection by XRT, continued monitoring revealed increased flux that turned out to be the largest X-ray flare ever detected during a GRB afterglow.  This giant X-ray flare was not accompanied by any detected emission in the BAT energy band.

\section{Observations}

In this paper, we are reporting observations and analysis with the XRT data set.  The XRT began taking data on this burst within 63 s after the BAT prompt emission trigger.  After the initial data taken in Image Mode, the spacecraft executed its usual sequence of modes and began taking Windowed Timing (WT) mode data.  Since the rate was initially low, the telescope then switched into Photon Counting (PC) mode.  Throughout the initial data segment, there was significant mode switching between PC mode and WT mode.  The initial data segment (data taken after the autonomous slew) contained 57.8 ks of data after screening.  Following this autonomous slew, follow-up observations were scheduled on GRB 050502B, which resulted in a total of 178 ks of data taken over a time period ranging from 63 s to 10.6 days after the BAT trigger.

\section{Analysis}

Data were reduced using the latest HEAsoft tools (version 6.0), including {\it{Swift}} software version 2.0, and the latest response (version 7.0) and ancillary response files (created using xrtmkarf) available in CALDB at the time of analysis.  Data were screened with standard parameters.  Data were also screened to eliminate time periods when the CCD temperature was warmer than -50$^\circ$ C.  When analyzing WT data, only grades 0--2 were included, and when using PC mode data, only grades 0--12 were included.  Source and background regions were specified independently for each data segment using the image from all data in that segment.  For PC mode data, the source region was a 30 pixel radius circle during the initial data segment when the source was very bright, and it was a 15 pixel radius circle for all subsequent data segments.  The background region was chosen in a source-free region of the image and was a circle with twice the radius of the source region in each data segment.  Background for WT mode data was found to be negligible.  The lightcurves were binned such that all bins contained $>8$ counts after background subtraction (actually, most bins had hundreds of counts).

Spectral models were fit to data using Xspec version 12.2.0.  Spectra were fit in the 0.2-5.5 keV energy range.  The energy range above 5.5 keV was excluded since some of the time regions, before and after the giant flare, had almost no counts above this energy.  In the interests of maintaining consistency in the analysis, a common upper bound (5.5 keV) was chosen for the energy range used in the fit.  A systematic error of 3\% was assigned throughout the energy range due to uncertainties in the response of the instrument.  The response below 0.6 keV is more uncertain since calibration is more difficult in this energy range, thus a systematic error of 5\% is assumed in the range 0.2--0.6 keV.  During fitting, $\chi^{2}$ statistics were used when appropriate (always with $>$20 photon/bin).  In the pre-flare time region when there were low counts ($\approx60$ counts), C-statistic was used, and the spectral data were binned to 1 photon/bin, which is more appropriate for C-statistic \citep{cas79, nou89, arn05}.  In general, the C-statistic provides better parameter estimation when there are few or no background counts and when source counts are low, so it was used to estimate parameters during the fitting.  Following the parameter estimation, $\chi^{2}$ statistics were then applied to these parameter values as an extra consistency check. 

During the pre-flare and flare time regions, mode switching led to collection of both PC mode data and WT mode data.  When the count rate was high ($\gtrsim0.5$ c/s), WT mode data were used, rather than PC mode data, in order to avoid effects due to pile-up on the detector.  This did not detract significantly from the data since there was nearly equivalent WT and PC mode coverage throughout the time period before and during the flare when the count rate was high.  WT mode data is free of significant pile-up effects throughout the entire range of flux reported in this paper, since pile-up does not begin to effect WT mode data until ~1 Crab \citep{bur05a}.  The PC and WT mode lightcurves before and after the flare were compared to one another to be sure that the data points with little or no pile-up were consistent with one another.

\section{Results}

\subsection{Lightcurve}

The overall lightcurve for GRB 050502B is shown in Figure 1.  Since the PC mode data were piled-up during the bright flare, the more reliable WT mode data were used for periods when the count rate was high ($\gtrsim0.5$ c s$^{-1}$).  However, it is worth pointing out that the pile-up corrected and non piled-up data in PC mode match the WT mode data during the flare, thus there should be no systematic shift in a lightcurve that contains both types of data.  The overall lightcurve has many features.  There is an obvious underlying decay curve.  Superimposed on this decay is a rapid and intense rate increase, beginning at ($345 \pm 30$) s.  For clarity, we will refer to this large rate increase as the ``giant flare'' throughout the remainder of the text.  There is also significant shorter timescale variability near the peak of the giant flare when viewed in the hard band (1.0--10.0 keV), as shown in Figure 2.  Following the giant flare, the underlying decay continues at a decay rate consistent with the decay rate before the giant flare.  After several hours ($>10^4$ s), two significant bumps in the X-ray emission occur consecutively.  At an undetermined time during, or after, the bumps, the underlying decay becomes steeper.
\clearpage
\begin{figure}
\plotone{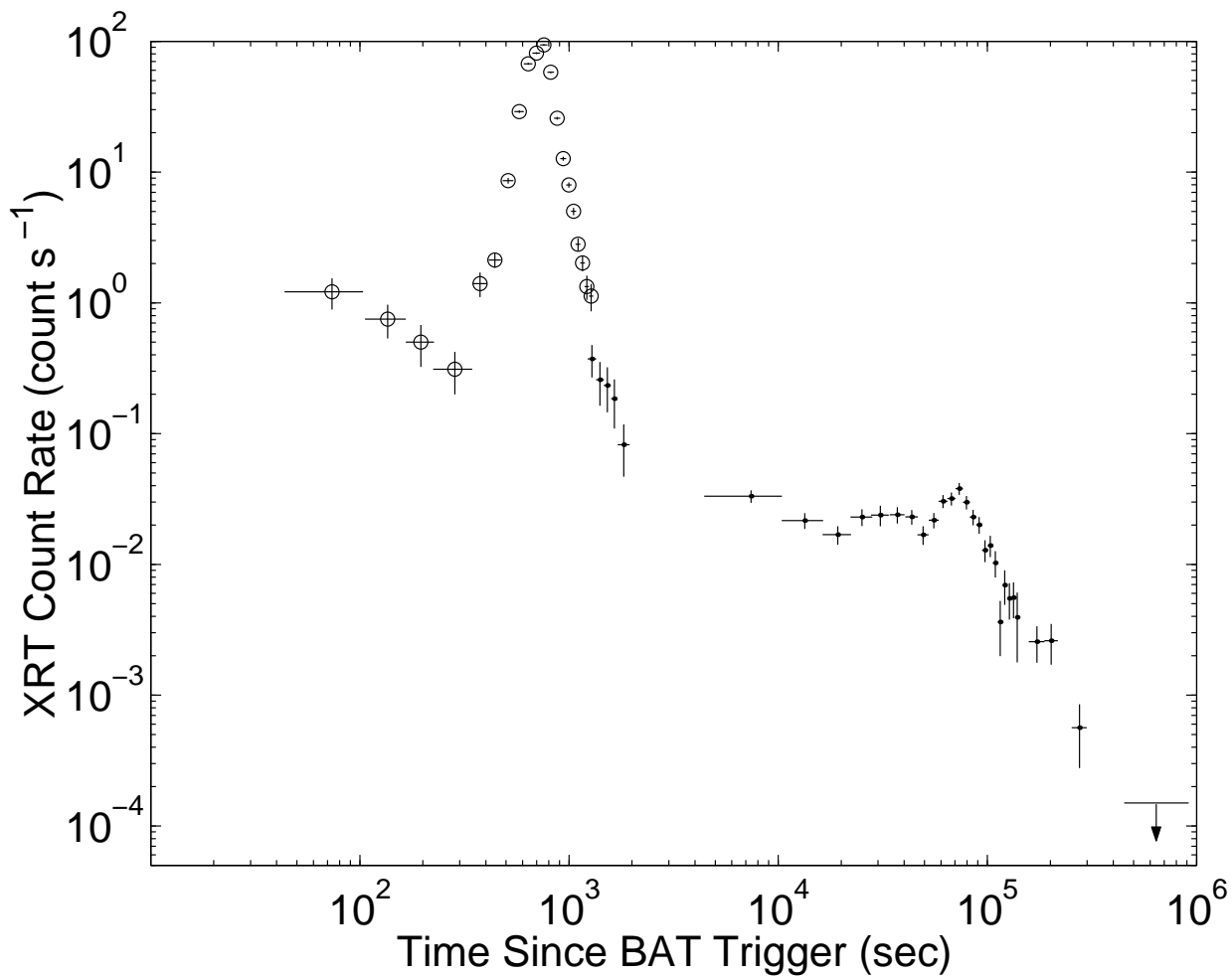}
\caption{Overall 0.2-10 keV band lightcurve of GRB 050502B.  Vertical error bars are $1\sigma$ statistical errors.  Horizontal error bars represent time bin size.  Open circle points represent WT mode data, and dots represent PC mode data.  The ``giant flare'' is the obvious $>500\times$ rate increase at ($345 \pm 30$) s.  There is also some bumpiness and/or flattening evident in the lightcurve at $\gtrsim10^4$ s, as well as an underlying decay below all of this activity.  Last datum is 90\% confidence level upper limit. \label{fig1}}
\end{figure}

\begin{figure}
\plotone{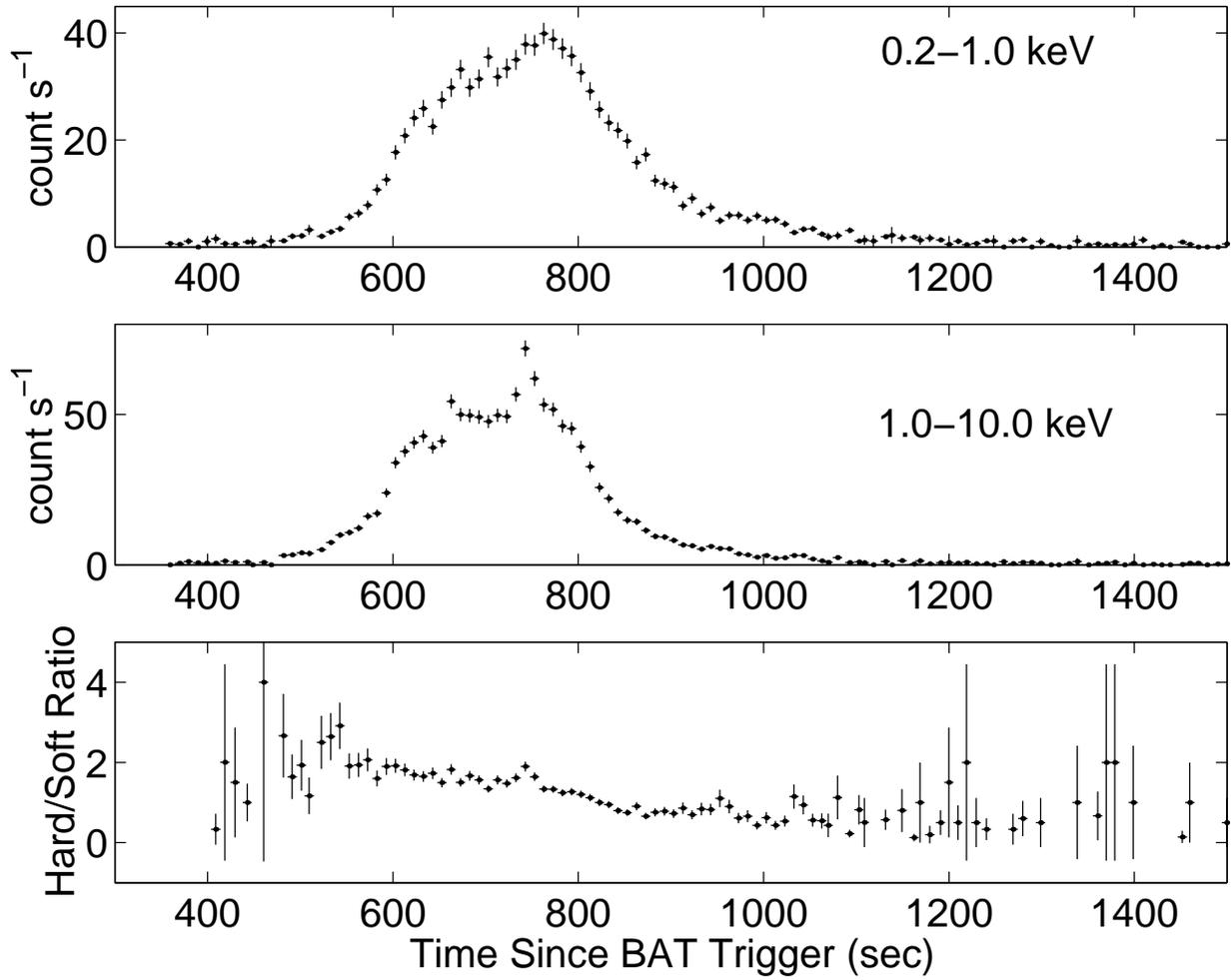}
\caption{Hard/soft band ratio of GRB 050502B.  Error bars are $1\sigma$ statistical errors.  \label{fig2}}
\end{figure}

\begin{figure}
\plotone{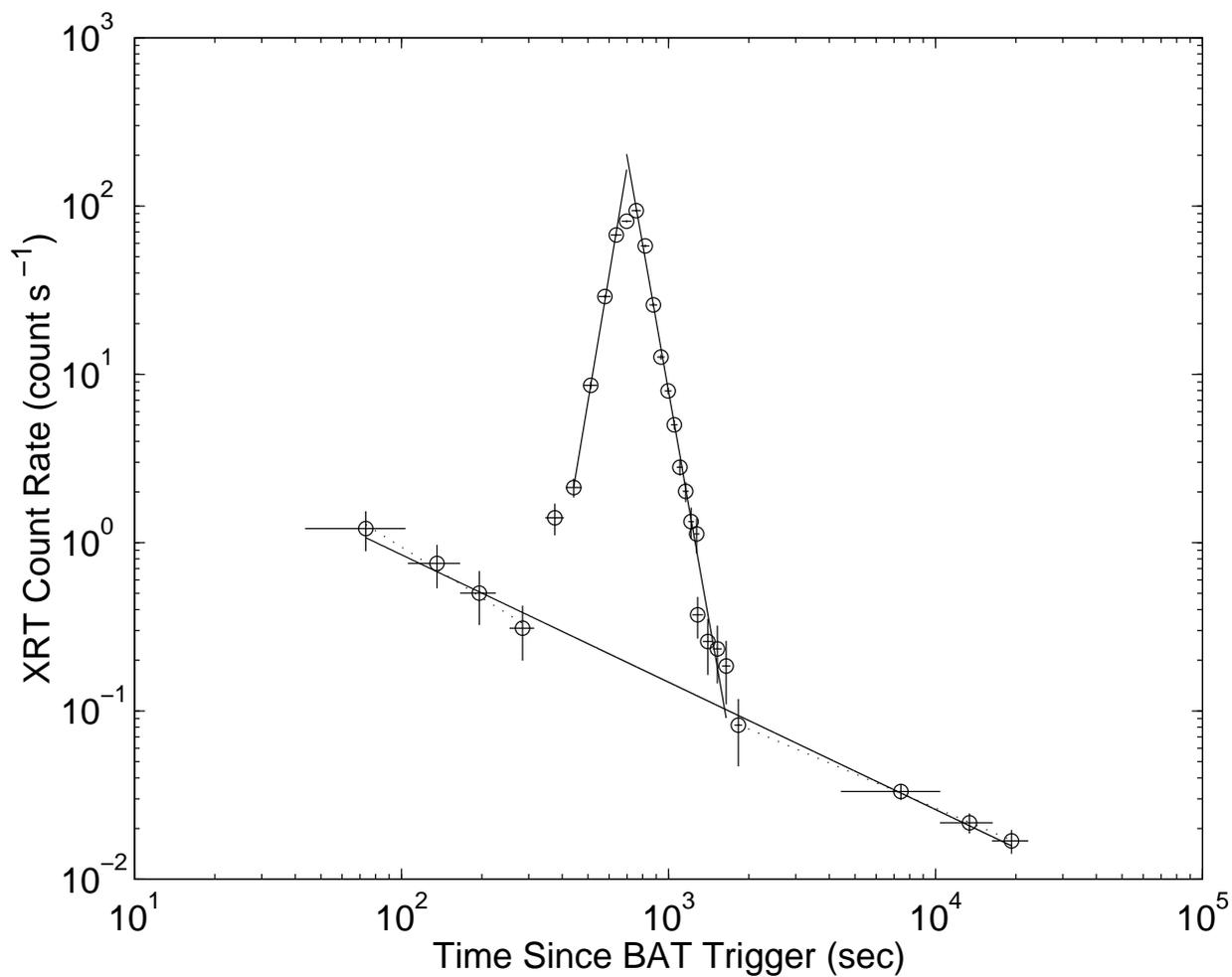}
\caption{Flare and fit to underlying decay curve of GRB 050502B, in the 0.2-10 keV energy band.  During the underlying decay, the solid line corresponds to a fit of all data points shown before and after flare.  The dotted lines, which are fits to the pre-flare and post-flare data independently, do not deviate significantly from fitting all underlying data simultaneously.  Vertical error bars are $1\sigma$ statistical errors.  Horizontal error bars represent the time bin size.  \label{fig3}}
\end{figure}
\clearpage
The underlying decay before and after the giant flare can be fit by a single power law that decays as $\sim t^{-\alpha}$, where $\alpha = 0.8 \pm 0.2$ (the error bar is dominated by systematic variations associated with the choice of flare start and stop times) and $t_{0}$ is taken as the time of the prompt burst.  As can be seen from Figure 3, fitting the pre-flare lightcurve data and the post-flare lightcurve data independently does not significantly improve the fit.  In this plot, the dashed lines represent fits to the pre-flare and post-flare data independently of one another (resulting in decay curves $\sim t^{-1.0 \pm 0.3}$ and $\sim t^{-0.7 \pm 0.3}$, respectively), while the solid line represents a fit of all pre-flare and post-flare data ($\sim t^{-0.8 \pm 0.2}$).  The lines in Figure 3 lie on top of one another, within error, indicating that there is no evidence for a change in the underlying decay after the giant flare, relative to that before the giant flare.  Aside from the flare itself, there is no evidence for any increase in the underlying decay flux relative to the pre-flare extrapolation of the decay (i.e. there is no evidence for afterglow shock re-energization near the time of the flare), but a factor of $\sim$2 increase of the post-flare flux relative to the pre-flare flux could be possible within the error bars of the data.

The giant flare begins to rise up above the underlying decay curve at 345 $\pm$ 30 s.  It rises to a sharp peak at 743 $\pm$ 10 s, but this appears to be on top of a broader peak that extends from 640 $\pm$ 20 s to 790 $\pm$ 20 s.  There is significant time structure within the peak of the giant flare itself, particularly in the harder X-ray lightcurve (see Figure 2).  The rise and decay of the giant flare can both be fit by simple power laws below the broad peak.  The flare rises rapidly as $\sim t^{9.5 \pm 2.1}$, and it decays rapidly as $\sim t^{-9.0 \pm 1.5}$.  If the underlying power law decay is subtracted from the giant flare data, then the rise is fit as $\sim t^{9.9 \pm 2.2}$, and the decay is fit as $\sim t^{-9.2 \pm 1.6}$.  The fits were all done using flare data outside of the 150 s region defining the peak and assuming a time origin at the burst trigger time.  Over the total 1360 s duration of the giant flare (837 s of good observation time), there were 23352 excess photons (in excess of both background and underlying decay curve extrapolation) collected in WT mode.  If one assumes that the 30 s duration hard X-ray rise at the top of the giant flare is a sub-flare, then the significance of only this short variation is 6.4 $\sigma$, with 376 photons detected in excess of the giant flare broad peak.

Following the decay of the giant flare, the lightcurve continues to decay at the pre-flare decay rate until $\sim10^{4}$ s.  Starting after (1.9 $\pm$ 0.3) $\times10^{4}$ s, there are two consecutive broad bumps in the lightcurve, or possibly a flattening combined with a bump.  The time frame for the first bump is estimated to be (1.9 $\pm$ 0.3) $\times10^{4}$ s to (4.9 $\pm$ 0.3) $\times10^{4}$ s, and the time frame for the second bump is taken as (4.9 $\pm$ 0.3) $\times10^{4}$ s to (1.1 $\pm$ 0.1) $\times10^{5}$ s.  The large error bar on the last data point is dominated by our inability to deconvolve the decay of the second bump from the faster decay of the underlying lightcurve after the second bump.  Given the time frames stated above, the signal-to-noise of the excess over the underlying decay curve for the first bump is 6.6 $\sigma$, and the signal-to-noise of the excess over the underlying decay for the second bump is 17 $\sigma$.

After the two late bumps (or possibly during the second bump), the slope of the lightcurve becomes steeper.  The last detection on the lightcurve and the upper limit shown in Figure 1 are significantly below the power law decay extrapolated from times prior to the late bumps.  The 90\% confidence level upper limit is a factor of 8 below the extrapolated curve.  It is difficult to estimate the time at which the steepening begins or the exact slope since we can not be certain of the contribution of the second late bump relative to the underlying decay curve itself.  If one fits the data above 1.1$\times10^{5}$ s with a power law decay, then the resulting decay index is $\alpha = 2.8 \pm 0.8$.

\subsection{Spectra}

\subsubsection{Hard/Soft Band ratio}

The hard/soft band ratio, as well as the lightcurves in the respective bands, are shown in Figure 2.  Three things are clear from this plot.  First, the hard/soft band ratio increases during the flare and subsequently decays as the flare decays.  We note the similarity to the spectral evolution of prompt GRB emission.  Second, the hard band shows significant short timescale variability, particularly near the peak of the giant flare (as discussed previously).  Third, the hard band begins to decay more quickly than the soft band.  Furthermore, the soft band has a shallower decay slope, relative to the slope of the flare rise.  It is clear from this plot that the hardness ratio increases at the onset of the flare, independent of any modelling, since this is merely a band ratio plot with no model dependence.

\subsubsection{Spectral Fits}

The spectra throughout the X-ray flaring and the underlying decay can be adequately fit by a simple power law with absorption, that looks like: 
\begin{equation}
f = C[e^{-N_{H}\sigma(E)}][\frac{E}{1 keV}]^{-\eta}
\end{equation}
where $N_{H}$ is the neutral Hydrogen column density with units $atoms/cm^{-2}$, $\sigma(E)$ is the energy dependent photoelectric absorption cross section \citep{mor83}, $\eta$ is the spectral photon index (note $\eta$ was chosen rather than the usual $\Gamma$ or $\alpha$ to avoid confusion with the bulk Lorentz factor or with the power law temporal decay index), and C is the normalization constant in units of photons cm$^{-2}$ s$^{-1}$ keV$^{-1}$.  These fits lead to $\chi^{2}/dof$ values that range from 0.67 to 1.21, as a function of the time used in the fit, except during the giant flare at which time the $\chi^{2}/dof$ reaches a value of 1.44 (328 dof) at the peak of the flare.  These fits lead to $N_{H}$ values that rise significantly during the onset of the flare and then decay after the flare; but outside of the giant flare time region, the $N_{H}$ values are reasonable.  Other models were explored and found to produce as good or superior fits, without the need for an apparent increase in the column density.

In general, the prompt emission from GRBs can be more realistically fit with Band functions \citep{ban93}, rather than simple power laws.  Since the giant X-ray flare could be due to the same mechanisms that produce the nominal prompt emission, it is reasonable to try to fit Band function or Band function-like models to the time periods when the flare was dominating the observed emission.  For the soft X-ray observations being analyzed here, nearly all of the photons are below any reasonable Band function cutoff energy.  As a result, spectral fits with a power law multiplied with an exponential cutoff are nearly equivalent to fits with a Band function, and these fits require fewer parameters and lead to better convergence on photon-limited data.  We have explored this by fitting several time periods during and around the giant flare with both absorbed Band functions and absorbed cutoff power laws.  We find that both of these models fit the flaring time regions significantly better than the simple power law models, whereas the absorbed power law models fit better during time periods when the flare was not dominating the emission.  The Band function models and the power laws with exponential cutoffs fit equally well during the flare, and the exponential cutoffs provide better convergence due to the reduced number of parameters to be fit.  At the peak of the flare, the cutoff power law fit the data with a $\chi^{2}/dof$ of 1.21.  The cutoff power law was compared to the simple power law using the ftest, resulting in an ftest probability of $\approx$3$\times10^{-14}$, supporting the improvement of the fit using the cutoff power law rather than the simple power law.  When this fact is coupled with the increasing column density values obtained using the simple power law and the prompt-emission-like timing properties of the giant flare, it is clear that the cutoff power law is more applicable to the flaring time region.  So for the remainder of the analysis, we have used the absorbed power law with exponential cutoff to fit time periods during the giant X-ray flare, and we have used an absorbed simple power law to fit all other time periods.  

The absorbed power law with exponential cutoff has the following form:
\begin{equation}
f = C[e^{-N_{H}\sigma(E)}][\frac{E}{1 keV}]^{-\eta}[e^{E/E_{cut}}],
\end{equation}
where $N_{H}$ is the neutral Hydrogen column density (as before), $\eta$ is the spectral photon index, $E_{cut}$ is the characteristic cutoff energy in units of keV, and C is the normalization constant in units of photons cm$^{-2}$ s$^{-1}$ keV$^{-1}$.  Since some models predict column density variations, we left $N_{H}$ free to vary throughout the fitting procedure, but we did set the lower bound of the $N_{H}$ to the measured Galactic value \citep{dic90}.

The results of applying the absorbed power law with an exponential cutoff model to the flare and an absorbed power law to the non-flare time regions are shown in Figure 4.  The value of $E_{cut}$ is not plotted in Figure 4 since it was consistent with a constant value of 2.5 $\pm$ 0.2 keV during the flare, with one exception.  During the final decay of the flare, $E_{cut}$ was large enough ($>$ 170 keV) to make the cutoff power law model effectively become a simple power law model with characteristics similar to the underlying decay.  There is significant variability of the soft X-ray spectral index during the giant flare, as can be seen in the first panel of Figure 4.  The data prior to the giant flare appear to have spectral fits that are consistent with the spectral fits after the flare.  During the giant flare, the spectrum hardens significantly; then, during the decay of the flare, it begins to soften back to a value consistent with the spectrum during the underlying lightcurve before and after the flare.  By the time the underlying emission is dominating the lightcurve again, the spectrum appears to have become soft again, as it was before the flare.  This softening during the decay of the flare is caused by the faster decay of the hard emission, as seen in Figure 2.  There is also a weak indication of some gradual hardening of the spectrum as the underlying flux fades.  All of the fit parameters displayed in figure 4 fit the data well; the ${\chi}^{2}/dof$ ranged from 0.62 to 1.21.  A sample spectral fit, calculated during the rising edge of the giant flare, is shown in Figure 5.
\clearpage
\begin{figure}
\plotone{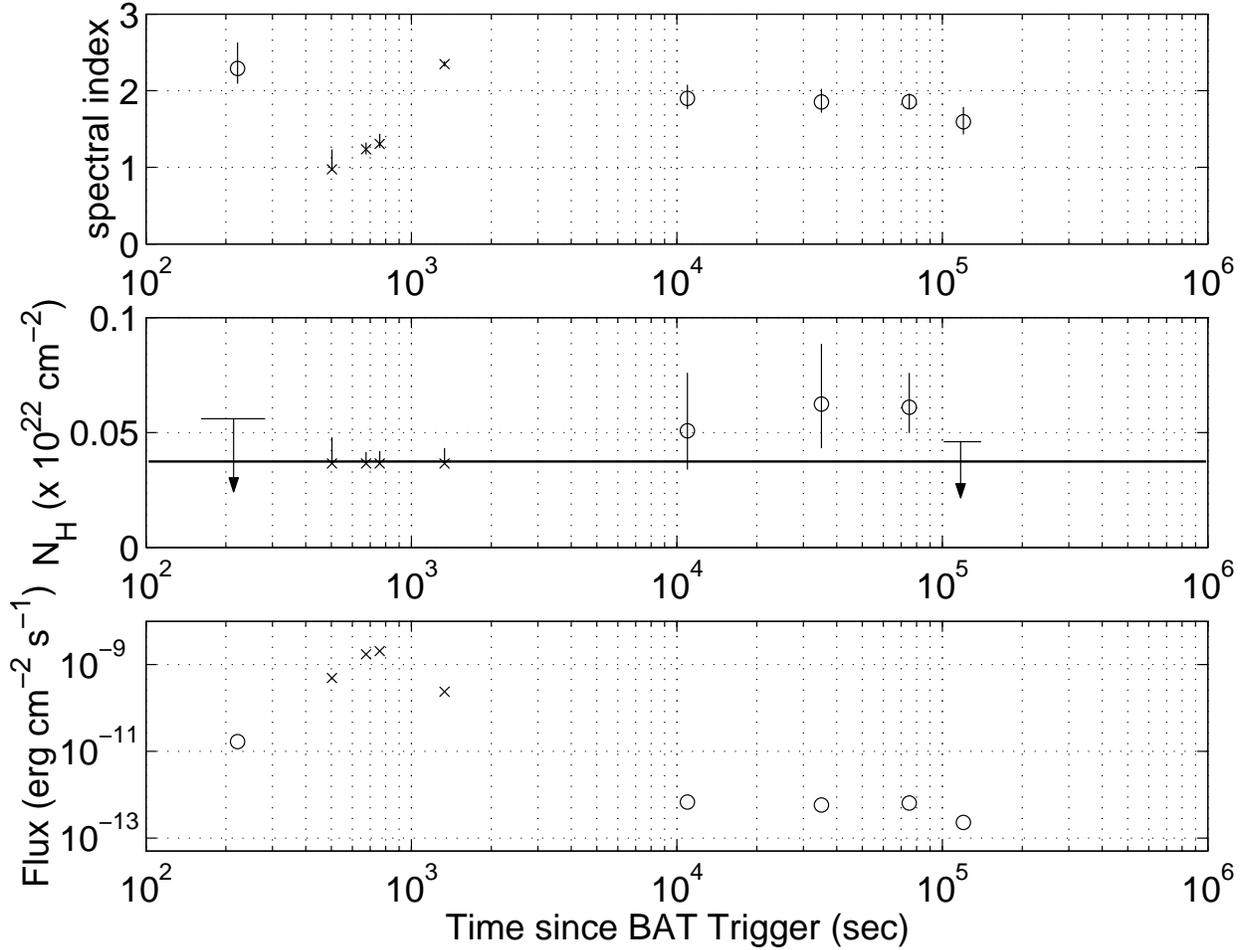}
\caption{Spectral parameters of GRB 050502B.  ``$\circ$'' data points (during non-flare time periods) represent parameter values fit using an absorbed simple power law model.  ``x'' data points (during giant flare) represent parameter values fit using an absorbed power law with an exponential cutoff.  Panel 1 shows an obvious hardening during the giant flare.  During the flare, the value for $E_{cut}$ was free, but it remained at 2.5 $\pm$ 0.2 keV, except during the transition from the flare decay to the underlying lightcurve dominated emission.  $N_{H}$, shown in second panel, was a free parameter, but it was required to be above the Galactic value \citep{dic90}, which is shown as a solid line across the plot; this is the reason for the uneven error bars during the flare.  Error bars are $1\sigma$ statistical errors, assuming two parameters of interest.  Upper limits are 90\% confidence level.  \label{fig4}}
\end{figure}

\begin{figure}
\plotone{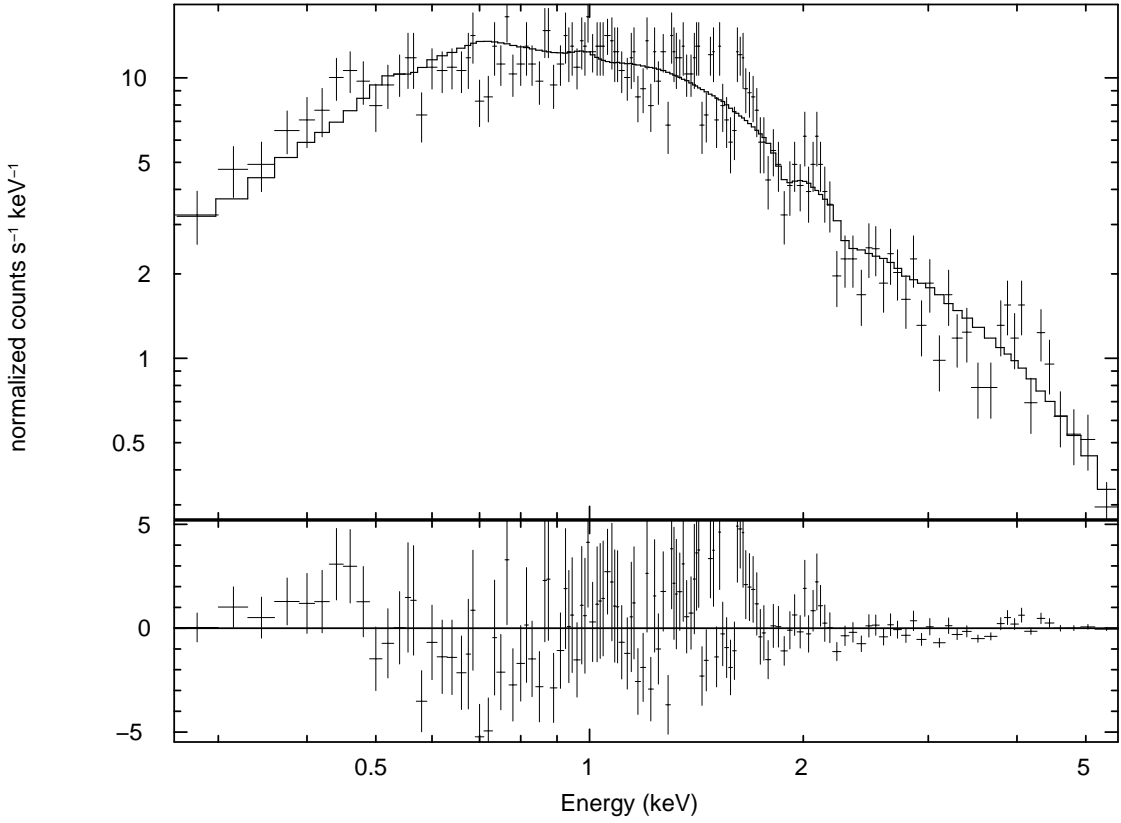}
\caption{An example of a spectral fit using the absorbed power law model with an exponential cutoff.  This fit used data during the rising edge of the giant X-ray flare.  Residuals of the fit are shown in the bottom panel.  \label{fig5}}
\end{figure}
\clearpage
\section{Energy of the Giant X-ray Flare}

During the time period of the giant X-ray flare, the total fluence in the 0.2--10.0 keV band was (1.2 $\pm$ 0.05) $\times 10^{-6}$ erg cm$^{-2}$.  After subtracting the underlying lightcurve data, the total fluence from the flare only is calculated to be (1.0 $\pm$ 0.05) $\times 10^{-6}$ erg cm$^{-2}$, in the same energy band.  This fluence actually exceeds the fluence measured during the prompt emission phase by BAT in higher energy bands, which was (8.0 $\pm$ 1.0) $\times10^{-7}$ erg cm$^{-2}$ in the 15--350 keV band \citep{cum05}.

\section{Discussion}

The fast variability timescales and the magnitude of the giant flare seem to favor an internal engine origin for the emission, as opposed to a mechanism associated with the afterglow.  Before looking at the supporting evidence for internal engine origin, let us first consider some afterglow-related possibilities.  Discontinuities in the jet environment can not explain the fast rise, $\sim t^{9.5 \pm 2.1}$ and small ${\delta}t_{rise}/t_{peak}\sim0.5$.  Rise and decay times associated with the forward shock encountering discontinuities are expected to be much longer than the times for these observations \citep{iok05}.  Explanations for the giant flare involving a refreshed shock are also ruled out since there is no evidence for a shift in the pre-flare versus post-flare lightcurve that would indicate energy injection into the external shock.  The pre-flare and post-flare lightcurves can be extrapolated to one another with near perfect overlap, indicating a continuous underlying afterglow extending from times well before the X-ray flare.  The giant flare is also steeper than one would expect if it was due to refreshing of the shock.  Explanations involving a reverse shock, and associated synchrotron and synchrotron self Compton emission, \citep{kob05, zha05} can be marginally accommodated by $\delta$t/t$\approx$1, however the large relative increase (factor of $\sim500\times$) of the giant flare over the afterglow and the faster variability at the peak of the flare makes these models less tenable in this case \citep{kob05,zha05}.  Furthermore, the synchrotron emission from a reverse shock model for this flare would require associated strong UV-optical flaring, which was not observed.

Interpreting this giant flare as the start of the afterglow phase \citep{pir05} is also inconsistent with the data.  It is apparent from the lightcurve and from the spectral index before and after the flare, that the afterglow began before the start of the X-ray flare.  The temporal power law decay index just before the flare matches the index just after the flare, as does the spectral index.  This favors independent mechanisms for production of the X-ray flare photons and the underlying afterglow photons.

Spectroscopically the afterglow and giant flare show interesting features.  Initially, we fit the whole evolution with an absorbed power law model, i.e. that expected from an afterglow created by a blast wave shock front. We found that this model provided a marginally good description of the data only if the absorbing column was left free to vary.  The resulting evolution tracks the flux evolution: the apparent $N_H$ is initially low, increases abruptly when the flare begins, and then
progressively decreases back to a value consistent with the pre-flare $N_{H}$.  \citet{laz02, laz03} predicted and discussed evolution of an X-ray absorbing column due to the progressive photoionization of the absorbing ions by the burst and afterglow UV and X-ray radiation.  The recombination timescale for any reasonable density of the ambient medium is longer than the observed times and therefore the X-ray column is expected to decrease in time.  Such behavior has been observed in a few GRBs with BeppoSAX \citep{laz02, fro04}.  For the giant flare from GRB 050502B, an {\it{increase}} of the absorbing column is required at the onset of the flare, if an absorbed power law is assumed, and even this does not provide a particularly good fit.  This is hard to interpret in terms of photoionization of the absorbing medium.  Small increases of the absorbing column can be accounted for if the external medium is clumped since the surface of the emitting fireball is increasing in time due to its deceleration.  However, an increase of the column by a factor of $\sim 10$ seems unlikely since the clumps will affect only a small portion of the visible fireball.

This apparently unrealistic $N_{H}$ increase, combined with the hypothesis that the giant flare was caused by the same mechanism as the nominal prompt emission (supported by the fast timing structure of the giant flare), motivated the fits to the data using a model that included an exponential cutoff.  This resulted in better fits to these data, and it resulted in more self consistent spectral parameters.  In particular, the $N_{H}$ no longer appeared to increase dramatically at the onset of the flare and was consistent with the Galactic value of $3.65\times10^{20}$ atoms cm$^{-2}$ \citep{dic90}.  The data does allow for the possibility that the $N_{H}$ was slightly higher after the flare, but it is certainly not required within the statistical and systematic errors that are present in the fitting procedure.  If the $N_{H}$ was slightly higher before and after the flare, with a decrease to Galactic values during the flare, this could be explained by photoionization, however, the final upper limit data point does not support this interpretation.  The simplest interpretation is that the $N_{H}$ was constant at approximately the Galactic value, with only slight fluctuations due to statistical and systematic errors.

As discussed above, we considered the possibility that the giant flare was due to late time internal shocks related to a late rebrightening of the inner engine.  In this case, the flare spectrum should be described with a Band function or an exponential cutoff power-law rather than a simple power-law, as is usually observed for typical prompt GRB emission.  Allowing for this extra degree of freedom, we find that a good fit can be obtained without the previously described dramatic evolution of the absorbing column.  The unabsorbed spectrum hardens sensibly during the flare, and the spectral shape is similar to that observed in X-ray flashes.  Following the giant flare, the spectral index returns to a value consistent with the softer afterglow spectrum observed before the flare.  This spectral evolution adds to the evidence discussed above for an internal shock origin of the giant flare.

The two bumps at $\gtrsim$10$^{4}$ s can be interpreted in a variety of ways.  Observationally, they are not well constrained.  They appear to be two late and broad flares with a subsequent increase in the afterglow decay rate (i.e. steepening of the lightcurve).  This increase in the decay rate could begin any time after the peak of the second bump, thus the interpretation of the timescales is difficult.  However, it is clear that there are two significant rate increases on top of the afterglow decay.  These two bumps could be continued internal engine activity, but the longer timescales and the spectral parameters do not require this interpretation.  The bumps could be caused by density fluctuations in the external medium through which the afterglow shock is propagating.  The bumps could also be due to energy injection into the afterglow shock front and/or an associated reverse shock.  This energy could come from material emitted from the internal engine that gets emitted at a later time than the initial prompt emission or that gets emitted with a slower speed.

The data shows no evidence of rebrightening immediately after the giant flare.  Evidence for rebrightening does not present itself until ${\gtrsim}10^4$~s.  If the giant flare is due to a delayed ejection episode from the central engine, this will eventually have to catch-up with the initially emitted ejecta and a brightening above the underlying afterglow by about a factor of 2 would be expected.  The time at which the catch-up is observed depends on the Lorentz factors of the two ejecta, on the energetics, and on the external medium properties.  In Figure 6, we explore the feasibility of this model using a contour plot of the observed time at which the re-energization would be expected to take place as a function of $\Gamma_1$ and $\Gamma_2$, the Lorentz factors of the material associated with the initial ejection episode and of the delayed one, respectively.  A total energy of $10^{52}$~erg for the initial ejection episode is assumed (relevant if the GRB lies at $z\sim1$ and the internal shock has a radiative efficiency of $\sim0.3$).  A uniform external medium with density $n=10$~cm$^{-3}$ is also assumed (assuming $n=1$~cm$^{-3}$ has only a marginal effect and does not impact conclusions).  The gray shaded area shows the region $10^4\le{t_{\rm{coll}}}\le10^5$~s, where the late time bumps of unknown origin are observed.  {\it{If}} one assumes that this model is correct, the plot shows two important things: 1) the second flow of ejecta started slower than the first one, and 2) it had a Lorentz factor $\Gamma_2\lesssim20$.
\clearpage
\begin{figure}
\plotone{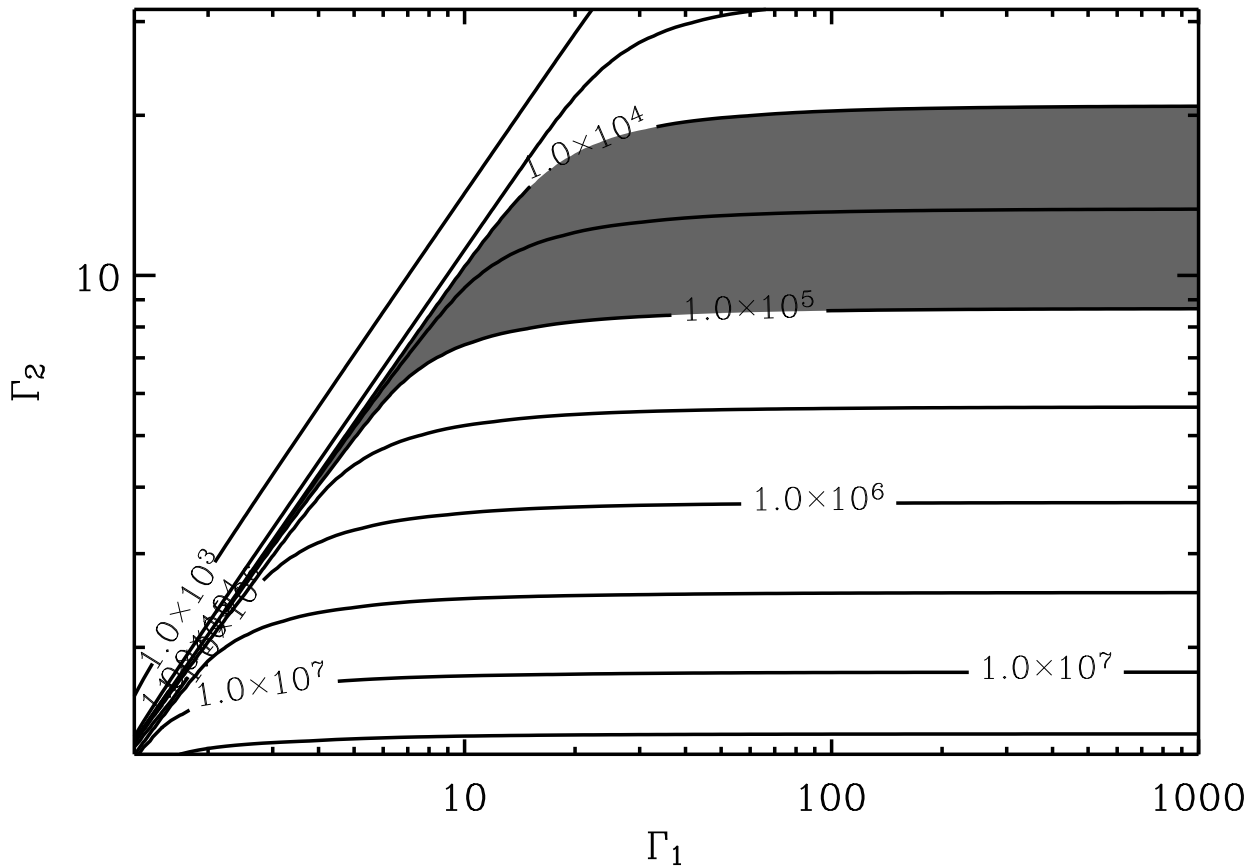}
\caption{Contour plot of the observed time of re-energization (in seconds) due to the collision between the first ejecta, which has initial Lorentz factor $\Gamma_1$ and hypothetically produces the initial prompt emission, with the delayed ejecta, which has Lorentz factor $\Gamma_2$ and hypothetically produces the giant X-ray flare emission. The shaded region shows the area consistent with the two late time bumps that may be caused by this collision. \label{fig6}}
\end{figure}
\clearpage
In the simplest version of synchrotron internal shocks, where the logarithmic dispersion of the Lorentz factor does not depend on the average $\Gamma$ of the flow, the peak frequency of the spectrum is inversely proportional to the Lorentz factor \citep{ghi00}.  In the model that attempts to explain the giant flare and the late time flux increase described in the paragraph above, we see the opposite behavior: the higher the Lorentz factor, the higher the peak frequency (since the larger $\Gamma_1$ is associated with the BAT detected prompt emission and the smaller $\Gamma_2$ is associated with the giant X-ray flare).  A way out of this inconsistency is to assume that the logarithmic dispersion of Lorentz factor is smaller when the average Lorentz factor is smaller.  This would produce smaller peak frequencies for smaller Lorentz factors.  Since the spectral properties of the flare are similar to X-ray flash spectra, it is tantalizing to infer that X-ray flashes might share this property.  Alternatively, one could attempt to explain the bumps at $>10^4 s$ as being due to even later internal engine activity.  More detailed studies to model the energy budget and the timing of sporadic and late internal engine emission will be necessary to discriminate between these two competing possibilities for the bumps in the lightcurve at times $>10^4 s$.

If one assumes the giant flare was due to a similar process as the initial prompt emission, the peak energy should be roughly proportional to ${L_{iso}}^{1/2}$, with significant spread around the expectation values, as shown for a sample of bursts by \citet{ghi05}.  Qualitatively, this is consistent with the data; the initial prompt emission relative to the giant flare emission peak luminosity is about 2 orders of magnitude higher and the peak energy decreases from ~80 keV to ~3 keV.  The downward shift of peak energy is possibly a bit more than expected by the simple scaling with ${L_{iso}}^{1/2}$, which could, in principle, be influenced by the relative Lorentz factors of the two ejection episodes.  However, the systematic errors due to the differing energy ranges of the initial prompt emission and the giant flare, as well as the measured spread in the peak energy to luminosity relationship, allow for order of magnitude variation.

In addition to flaring, a steepening was observed in the lightcurve beyond ${\sim}1.1\times10^{5}$ s.  One obvious interpretation of this is that a jet break has been observed.  A jet break is a steepening of the lightcurve decay that occurs when the relativistic beaming angle, which is increasing as the fireball Doppler factor is decreasing, begins to exceed the physical opening angle of the jet \citep{rho99}.  One can calculate the jet opening angle as a function of the jet break time, assuming a fireball model expanding into a constant density interstellar medium.  This leads to $\theta_0 = 7.8^{\circ} {t_{5}}^{3/8}{E_{52}}^{-1/8}{n_{ism}}^{1/8}[(1+z)/2]^{-3/8}$, where $t_{5}$ is jet break time in units of $10^{5}$ s, $E_{52}$ is isotropic energy in units of $10^{52}$ erg, $n_{ism}$ is the interstellar medium density with units $cm^{-3}$, and z is the redshift.  Assuming typical parameters and a jet break time of ${\sim}1.1\times10^{5}$ s leads to an opening angle of $\sim8.1^{\circ}$, which is within the typical range of bursts observed and studied with known redshifts \citep{blo03}.  Alternatively, this steeper slope could be interpreted as the end of a period of continuous energy injection that was feeding the afterglow shock, thus creating a shallower decay slope followed by a steepening when the continuous energy injection phase ends.  However, the work of \citet{zha05} suggests that this type of break would lead to a steepening of the lightcurve decay index to $\sim1.2$, whereas the data in this case suggest a steepening to a decay index of $\approx2.8\pm0.8$, which is more indicative of a jet break.  While the model in which the break is caused by an end to continuous energy injection is not ruled out here, it seems as though a jet break is the more likely explanation for the steepening in the lightcurve after ${\sim}1.1\times10^{5}$ s.  The data from this burst, as well as other flaring bursts found in the {\it{Swift}} XRT data, should allow much more rigorous studies of the various models and their predicted temporal and spectral parameters.

It is also possible, within the error bars of the temporal power law decay indices, that the underlying afterglow decay index was steeper ($\alpha\sim1.0$) throughout the whole curve and that the entire lightcurve, between the start of the giant flare and the final data point on Figure 1, was due to internal engine flaring and energy injection.  This scenario seems less likely since the features of the lightcurve and spectra before and after the giant flare match each other well and since it $\it{requires}$ even later internal engine activity.  This scenario is mentioned since it is possible within the error bars of the data; however, the data certainly do not require or suggest this interpretation.

\section{Conclusions}

The complex lightcurve of GRB 050502B has many interesting features.  The most interesting of these is a giant flare with a fast rise/decay that began at 345 $\pm$ 30 s, well after the nominal 17.5 $\pm$ 0.2 s prompt GRB phase.  This is the largest X-ray flare ever detected after the apparent cessation of prompt emission.  After compiling all of the evidence, we come to the conclusion that the simplest explanation for this flare is continued activity in the internal engine of the GRB, not associated with the afterglow external shock.  This evidence includes: 1) The temporal decay index before and after the flare are identical, indicating that the afterglow had already begun before the flare, 2) the rise time and decay time of the flare are very fast, thus the flare is difficult (although not impossible) to explain with mechanisms associated with the external shock, 3) there is even faster time structure near the peak of the flare in the band above 1 keV, 4) the spectra during the giant flare are represented better by a Band function or cutoff power law model, rather than a simple power law, 5) the spectra before and after the flare are consistent with an afterglow that has already begun before the flare and continues with approximately the same spectral index after the flare, whereas the spectra during the flare is significantly harder, 6) based on recent data there appear to be many other similar flares, although not as intense as this one, with even faster time profiles (suggesting extended internal engine activity) in up to $\sim$1/2 of the {\it{Swift}} detected GRBs.

The two late bumps in the lightcurve (at $> 10^{4}$ s) are not as well constrained so conclusions are not as firm.  They could be due to either more internal engine activity, such as that which created the giant flare, or they could be due to another process associated with the afterglow shock.  If one or both of these bumps is due to re-energization of the afterglow shock due to the same internal engine ejection episode that created the giant flare at $\sim$345 s, then this ejecta must have been emitted with a Lorentz factor $\Gamma_2\lesssim20$, and it must have been travelling slower than the primary ejecta that created the BAT detected prompt emission.  However, the limits on $\Gamma_2$ are obviously dependent on the presumption that the bumps at $>10^4 s$ are due to re-energization of the afterglow shock by the same ejecta responsible for the giant flare.  Although this presumption is consistent with the data, it is not the only possible explanation, as stated earlier.  We can also conclude that there was a steepening of the lightcurve at some time during or after the late bumps (i.e. after $\sim5\times10^{4}$ s).  This steepening could have been a jet break, or it could have been the end of a phase of continuous energy injection into the afterglow shock front.

This GRB, and the associated X-ray flare, can be most easily explained within the framework of the standard GRB fireball model \citep{mes97}, provided that there is some mechanism to feed the internal engine activity for extended time periods.  Although "Type I" collapsar models with prompt black hole formation cannot explain late time internal engine activity \citep{mac01}, the fallback of material onto the central black hole after a stellar collapse could last for long time periods \citep{woo93, mac01} and lead to late internal engine activity, albeit with significantly reduced luminosity.  Continued energy release due to dynamics of a magnetized disk around a black hole, as described by \citet{kat97}, and/or continued and sporadic emission due to fragmentation and subsequent accretion during the collapse of a rapidly rotating stellar core, as described by \citet{kin05}, could both explain observations of extended production of internal shocks.

In the time period since the analysis of the data from this GRB, {\it{Swift}} has detected several more GRBs with X-ray flares \citep{nou05}.  In the near future, it will be possible to use samples of many GRBs with X-ray flares to test models of long timescale prompt emission. 

\acknowledgments


\begin{thebibliography}{}

\bibitem[Arnaud et~al.\ (2005)]{arn05} Arnaud, K., Dorman, B., Gordon, C. 2005, Xspec User's Guide Version 12.2

\bibitem[Band et~al.\ (1993)]{ban93} Band, D., Matteson, J., Ford, L., et~al.\ 1993, \apj, 413, 281

\bibitem[Barthelmy et~al.\ (2004)]{bar04} Barthelmy, S., et~al.\ 2004, Proc. of SPIE, 5165, 175

\bibitem[Bloom et~al.\ (2003)]{blo03} Bloom, J.S., Frail, D.A., \& Kulkarni, S.R. 2003, \apj, 594, 674

\bibitem[Burrows et~al.\ (2005a)]{bur05a} Burrows, D.N., Hill, J.E., Nousek, J.A., et~al.\ 2005a, \ssr, in press; astro-ph/0508071

\bibitem[Burrows et~al.\ (2005b)]{bur05b} Burrows, D.N., Romano, P., Falcone, A., et~al.\ 2005b, Science, Vol. 309, Issue 5742, 1833

\bibitem[Cash (1979)]{cas79} Cash, W. 1979, \apj, 228, 939

\bibitem[Cenko et~al.\ (2005)]{cen05} Cenko, S.B., Fox, D.B., Rich, J., et~al.\ 2005, GCN Circular 3358

\bibitem[Connaughton (2002)]{con02} Connaughton, V. 2002, \apj, 567, 1028

\bibitem[Cummings et~al.\ (2005)]{cum05} Cummings, J., Barbier, L., Barthelmy, S., et~al.\ 2005, GCN Circular 3339

\bibitem[Dickey \& Lockman (1990)]{dic90} Dickey, J.M. \& Lockman, F.J. 1990, Ann. Rev. Astron. \& Astrophys., 28, 215

\bibitem[Falcone et~al.\ (2005)]{fal05} Falcone, A., Burrows, D.N., Chester, M., et~al.\ 2005, GCN Circular 3330

\bibitem[Frontera et~al.\ (2004)]{fro04} Frontera, F., et~al.\ 2004, \apj, 614, 301

\bibitem[Galli \& Piro (2005)]{gal05} Galli, A. \& Piro, L. 2005, submitted to Astron. \& Astrophys.; astro-ph/0510852

\bibitem[Gehrels et~al.\ (2004)]{geh04} Gehrels, N., Chincarini, G., Giommi, P., et~al.\ 2004, \apj, 611, 1005

\bibitem[Ghirlanda et~al.\ (2005)]{ghi05} Ghirlanda, G., Ghisellini, G., Firmani, C., Celotti, A., Bosnjak, Z. 2005, \mnras, 360, L45

\bibitem[Ghisellini et~al.\ (2000)]{ghi00} Ghisellini, G., Celotti, A., Lazzati, D. 2000, \mnras, 313, L1

\bibitem[in't Zand et al.\ (2003)]{int03} in't Zand, J.J.M., Heise, J., Kippen, R.M., et al. 2003, Proceedings of 3rd Rome Workshop, ASP Vol. 312, 18; astro-ph/0305361

\bibitem[Ioka et~al.\ (2005)]{iok05} Ioka, K., Kobayashi, S., \& Zhang, B. 2005, \apj, accepted; astro-ph/0409376

\bibitem[Katz (1997)]{kat97} Katz, J.I. 1997, \apj, 490, 633

\bibitem[King et~al.\ (2005)]{kin05} King, A., O'Brien, P.T., Goad, M.R., et~al.\ 2005, \apjl, accepted; astro-ph/0508126

\bibitem[Kobayashi et~al.\ (2005)]{kob05} Kobayashi, S., Zhang, B., M\'{e}sz\'{a}ros, P., \& Burrows, D. 2005, \apjl, submitted; astro-ph/0506157

\bibitem[Lazzati \& Perna(2002)]{laz02} Lazzati, D., \& Perna, R. 2002, \mnras, 330, 383 

\bibitem[Lazzati \&  Perna(2003)]{laz03} Lazzati, D., \& Perna, R. 2003, \mnras, 340, 694

\bibitem[MacFadyen et al (2001)]{mac01} MacFadyen, A.I., Woosley, S.E., \& Heger, A. 2001, \apj, 550, 410

\bibitem[Meegan et~al.\ (1996)]{mee96} Meegan, C.A., et~al.\ 1996, \apjs, 106, 65

\bibitem[M\'{e}sz\'{a}ros \& Rees (1997)]{mes97} M\'{e}sz\'{a}ros, P. \& Rees, M.J. 1997, \apj, 476, 232

\bibitem[M\'{e}sz\'{a}ros (2002)]{mes02} M\'{e}sz\'{a}ros, P. 2002, Ann. Rev. Astron. \& Astrophys.,40, 137

\bibitem[Morrison and McCammon (1983)]{mor83} Morrison, R., \& McCammon, D. 1983, \apj,270, 119

\bibitem[Nousek \& Shue (1989)]{nou89} Nousek, J.A. \& Shue, D.R. 1989, \apj,342, 1207

\bibitem[Nousek et~al.\ (2005)]{nou05} Nousek, J.A., Kouveliotou, C., Grupe, D., et~al.\ 2005, \apj, submitted; astro-ph/0508332

\bibitem[Pagani et~al.\ (2005)]{pag05} Pagani, C., Falcone, A., Burrows, D.N., et~al.\ 2005, GCN Circular 3333

\bibitem[Piro et~al.\ (2005)]{pir05} Piro, L., et~al.\ 2005, \apj, 623, 314

\bibitem[Piran (2005)]{piran05} Piran, T. 2005, Rev. Mod. Phys., 76, 1143

\bibitem[Rhoads (1999)]{rho99} Rhoads, J.E. 1999, \apj, 525, 737

\bibitem[Romano et~al.\ (2005)]{romano05} Romano, P., et~al.\ 2005, in preparation

\bibitem[Roming et~al.\ (2005)]{rom05} Roming, P., et~al.\ 2005, \ssr, in press; astro-ph/0507413

\bibitem[Van Paradijs et~al.\ (2000)]{van00} van Paradijs, J., Kouveliotou, C., \& Wijers, R.A.M.J. 2000, Ann. Rev. Astron. \& Astrophys.,38, 379

\bibitem[Watson et~al.\ (2003)]{wat03} Watson, D., Reeves, J.N., Hjorth, J., Jakobsson, P., \& Pederson, K. 2003, \apjl, 595, L29

\bibitem[Woosley (1993)]{woo93} Woosley, S.E. 1993, \apj, 405, 273

\bibitem[Zhang \& M\'{e}sz\'{a}ros (2004)]{zha04} Zhang, B. \& M\'{e}sz\'{a}ros, P. 2004, Int. Journ. of Mod. Phys. A, 19, 2385

\bibitem[Zhang et~al.\ (2005)]{zha05} Zhang, B., Fan, Y.Z., Dyks, J., et al. 2005, \apj, submitted; astro-ph/0508321


\end{thebibliography}
\end{document}